\documentclass[british,amssymb,aps,prd,twocolumn,amsmath,floats,floatfix,superscriptaddress,nofootinbib,%
               showkeys,showpacs,groupedaddress]{revtex4-1}
\usepackage{amssymb}
\usepackage{amsmath}
\usepackage{verbatim}
\usepackage{mathrsfs}
\usepackage{amsfonts}
\usepackage{latexsym}
\usepackage{epsfig}
\usepackage{color}
\usepackage[usenames,dvipsnames]{xcolor}
\usepackage{graphicx,subfigure}
\usepackage{units}
\usepackage{envmath}
\usepackage{natbib}
\usepackage{ctable}
\usepackage{soul}
\usepackage[unicode=true,pdfusetitle,
 bookmarks=true,bookmarksnumbered=false,bookmarksopen=false,
 breaklinks=false,pdfborder={0 0 1},backref=false,colorlinks=false]{hyperref}
\usepackage{mathtools}
\begin{document}
\definecolor{orange}{rgb}{0.9,0.45,0}
\def\CovDev{D}
\def\Res{{\mathcal R}}
\def\Gammaflat{\hat \Gamma}
\def\metricflat{\hat \gamma}
\def\Dflat{\hat {\mathcal D}}
\def\part_n{\partial_\perp}
\def\Lie{\mathcal{L}}
\def\A{\mathcal{X}}
\def\Aphi{\A_{\phi}}
\def\hAphi{\hat{\A}_{\phi}}
\def\E{\mathcal{E}}
\def\Ham{\mathcal{H}}
\def\M{\mathcal{M}}
\def\R{\mathcal{R}}
\def\p{\partial}
\def\hg{\hat{\gamma}}
\def\hA{\hat{A}}
\def\hD{\hat{D}}
\def\hE{\hat{E}}
\def\hR{\hat{R}}
\def\hcA{\hat{\mathcal{A}}}
\def\hDelt{\hat{\triangle}}
\def\na{\nabla}
\def\dif{{\rm{d}}}
\def\non{\nonumber}
\newcommand{\erf}{\textrm{erf}}
\renewcommand{\t}{\times}
\long\def\symbolfootnote[#1]#2{\begingroup%
\def\thefootnote{\fnsymbol{footnote}}\footnote[#1]{#2}\endgroup}
\title{Thermodynamics of Deformed AdS-Schwarzschild Black Hole}
\author{Mohammad Reza Khosravipoor}
\email{m\_khosarvipoor@sbu.ac.ir}
\author{Mehrdad Farhoudi}
\email{m-farhoudi@sbu.ac.ir}
 \affiliation{Department of Physics,
              Shahid Beheshti University, Evin, Tehran 19839, Iran}

\date{November 13, 2023}

\begin{abstract}
\noindent
 By implementing the gravitational decoupling method, we
find the deformed AdS-Schwarzschild black hole solution when there
is also an additional gravitational source, which obeys the weak
energy condition. We also deliberately choose its energy density
to be a certain monotonic function consistent with the
constraints. In the method, there is a positive parameter that can
adjust the strength of the effects of the geometric deformations
on the background geometry, which we refer to as a deformation
parameter. The condition of having an event horizon limits the
value of the deformation parameter to an upper bound. After
deriving various thermodynamic quantities as a function of the
event horizon radius, we mostly focus on the effects of the
deformation parameter on the horizon structure, the thermodynamics
of the solution and the temperature of the Hawking-Page phase
transition. The results show that with the increase of the
deformation parameter: the minimum horizon radius required for a
black hole to have local thermodynamic equilibrium and the minimum
temperature below which there is no black hole decrease, and the
horizon radius of the phase transition and the temperature of the
first-order Hawking-Page phase transition increase. Furthermore,
when the deformation parameter vanishes, the obtained
thermodynamic behavior of the black hole is consistent with that
stated in the literature.
\end{abstract}

\maketitle

\section{Introduction}
Before the 1970s, there was no compelling reason to study the
thermodynamics of black holes, until Hawking's area theorem
changed this view~\cite{HAT}. Then, Bekenstein linked Hawking's
area theorem with the second law of thermodynamics by assigning
entropy to black holes~\cite{BE}. Using such similarity and
assuming the event horizon of black holes as a Killing horizon,
the four laws of black hole mechanics were formulated~\cite{FLOB}.
However, with the classical view that black holes absorb all
matter and energy and emit no radiation, attributing temperature
and entropy to black holes was questionable. But by using quantum
effects, Hawking showed that black holes have radiation with the
spectrum of a black body with a certain temperature~\cite{HTOB}.
After the formulation of the standard laws of thermodynamics of
black holes, several interpretations and investigations were
carried out in the thermodynamics committee about black holes,
such as entropy and temperature, the first law of thermodynamics,
and relations between extensive and intensive quantities, see,
e.g., Refs.~\cite{waldTB,Carlip} and references therein.

The thermodynamic laws of ordinary materials are associated with a
pressure-volume term, however those for black holes do~not contain
such a term. To overcome this inconsistency, an idea based on the
immersion of black holes in a background with a negative
cosmological constant was presented~\cite{Black hole chemistry}.
In this view, pressure can be considered equal to a negative
cosmological constant. In this regard, in the context of black
hole thermodynamics, black holes with asymptotic anti-de Sitter
(AdS) are more interesting. The idea of using the cosmological
constant as the thermodynamic pressure~\cite{Tidea,Bidea}
generalized the first law of black hole thermodynamics and formed
an extended phase-space, see, e.g.,
Refs.~\cite{Farhangkhah,Bhattacharya}. Hence, in the presence of
the cosmological constant, the mass of a black hole does~not
represent the internal energy of the thermodynamic system and is
interpreted as the gravitational version of the
enthalpy~\cite{Enthalpy by kastor}.

The new perspective of black hole thermodynamics based on the new
interpretation of the cosmological constant and black hole mass
has led to new phenomena associated with black holes. Accordingly,
topics such as thermodynamic equilibrium and phase transition can
be addressed. One of the most important thermodynamic behaviors is
the phase transition between sufficiently large (compared to the
AdS radius) black holes and an environment full of special
radiation. This is known as the Hawking-Page phase
transition~\cite{Hptransition}, which is considered as a
confinement/deconfinement phase transition in boundary conformal
field theory~\cite{CFTWitten}.

Furthermore, black hole thermodynamics has wide range of
applications compared to the Einstein gravity. Even, the
investigation of more complex behaviors in the issue of the
thermodynamic phase transition of black holes in modified
gravitational theories has received much attention, see, e.g.,
Refs.~\cite{Lovelock,Born-Infeld}. In general, finding a black
hole as a solution in a gravitational theory leads to the
investigation of thermodynamic behavior from the researchers'
point of view. Subsequently, more studies have been conducted and
other novel phenomena and phase structures have been observed,
see, e.g., Refs.~\cite{Rotating bh,P-v cp,triple point,Isolated
cp,third lovelock,dRGT}.

The non-linear nature of equations in gravitational theories leads
to approximate methods and, of course, innovative proposals for
finding analytical solutions, which is always an important task.
In this regard, one of the attractive approach for searching and
analyzing solutions of the gravitational equations is the
gravitational decoupling (GD) method, which serves as a useful
tool, see, e.g., Refs.~\cite{Ovalle2017,extended MGD,Ovalle2023}
and references therein. Through this method, the known solutions
of the standard gravitational action can be extended (with some
minimal set of requirements) to additional sources and the domain
of modified gravitational theories. The implementation of the GD
method makes it possible to decouple the gravitational equations
of sources into two parts, one for the standard field equations
and one for an additional gravitational source. This method has
been applied in the context of the Randall-Sundrum
brane-world~\cite{compact star,Tolman IV} and has also been
extended to investigate in other gravitational issues, including
new black hole solutions, see, e.g.,
Refs.~\cite{Ovalle2023,minimal
geo,Silva2018,Ovalle2018,Rincon2019,Rincon2020,Rocha2020,Ovalle2021,axially
sym,Mahapatra,Avalos2023,Avalos-etal,Meng-etal,Tello}. In the
present work, while employing the GD method, we intend to extend
the AdS-Schwarzschild vacuum solution in the presence of a generic
gravitational source, which satisfies the weak energy condition.
Thereafter, we try to find possible deformed AdS-Schwarzschild
black hole solution and then investigate its thermodynamic
properties and Hawking-Page phase transition while varying the
relevant parameters.

The work is organized as follows. In Sec.~II, while briefly
reviewing and implementing the GD method, we introduce the desired
gravitational action. In Sec.~III, we derive the deformed
AdS-Schwarzschild black hole. In Sec.~IV, we scrutinize the
structure of the horizon and calculate the thermodynamic
quantities in the extended phase-space to investigate the horizon
structure, the thermodynamics of the solution and the temperature
of the Hawking-Page phase transition. Finally, we summarize the
results in Sec.~V.

\section{Modified Gravitational Action and Decoupling Field Equations}
We consider the following action in four dimensions in which an
additional general Lagrangian term is added to the standard
Einstein gravity with the cosmological constant, namely
\begin{equation}\label{main action}
    A=\int{d^4 x\,
    \sqrt{-g}\,\left(\frac{R-2\Lambda}{2\kappa}+L_m+L_{\rm X}\right)},
\end{equation}
where $g$ is the determinant of the metric, $R$ is the Ricci
scalar, $\Lambda$ is the cosmological constant, $L_m $ is the
usual matter Lagrangian and $L_{\rm X} $ represents Lagrangian for
any other matter or new gravitational sector beyond general
relativity, e.g. Lovelock gravity and/or new other
scalar/vector/tensor field(s). Meanwhile, $\kappa=8\pi G_{\rm
N}/c^4$, however through the work, we use the natural units
$\hbar=1=c$, and also employ the $(-,+,+,+)$ signature. The
variation of the action with respect to the metric gives the field
equations
\begin{equation}\label{enesteineq}
    G_{\mu \nu}+\Lambda g_{\mu\nu} =\kappa\, T^{(\rm tot)}_{\mu\nu},
\end{equation}
where $G_{\mu\nu}$ is the Einstein tensor and $T^{(\rm
tot)}_{\mu\nu}$ represents the total symmetric energy-momentum
tensor as
\begin{equation}\label{ems}
    T^{(\rm tot)}_{\mu\nu}=T^{(\rm m)}_{\mu\nu}+T^{(\rm X)}_{\mu\nu},
\end{equation}
wherein
\begin{equation}\label{line elemets3}
T^{(\rm m)}_{\mu\nu}=-\frac{2}{\sqrt{-g}}\frac{\delta(\sqrt{-g}\,
L_m)}{\delta g^{\mu\nu}}
\end{equation}
and
\begin{equation}\label{line elemets4}
    T^{(\rm X)}_{\mu\nu}=-\frac{2}{\sqrt{-g}}\frac{\delta(\sqrt{-g}\, L_{\rm X})}{\delta
        g^{\mu\nu}}.
\end{equation}

As an ansatz, we demand the solution to be the spherically
symmetric and static spacetime, namely
\begin{equation}\label{line elemets1}
    ds^2=-e^{\nu(r)}dt^2+e^{\mu(r)}dr^2+r^2(d\theta^2+sin^2{\theta}\,
    d\phi^2),
\end{equation}
in the spherical coordinates. Hence, metric~\eqref{line elemets1}
must satisfy the field Eqs.~\eqref{enesteineq}, namely
\begin{equation}\label{00}
\kappa\left(T^{(\rm m)}{}_0^{0}+T^{(\rm X)}{}_0^{0}\right)=\Lambda
+e^{-\mu } \left(\frac{1}{r^2}-\frac{\mu
'}{r}\right)-\frac{1}{r^2},
\end{equation}
\begin{equation}\label{11}
\kappa\left(T^{(\rm m)}{}_1^{1}+T^{(\rm X)}{}_1^{1}\right)=\Lambda
+e^{-\mu } \left(\frac{1}{r^2}+\frac{\nu
'}{r}\right)-\frac{1}{r^2},
\end{equation}
\begin{eqnarray}\label{22}
\kappa\left(T^{(\rm m)}{}_2^{2}+T^{(\rm X)}{}_2^{2}\right)=\Lambda
 &+&\frac{1}{4} e^{-\mu }\Big[ -
 \mu '\nu '+2 \nu ''\cr
 &+&\nu '^2+\frac{2 \left(\nu '-\mu
'\right)}{r}\Big],
\end{eqnarray}
where prime denotes derivative with respect to $r$ and $ T^{(\rm
tot)}{}_3^{3}=T^{(\rm tot)}{}_2^{2}$ due to the spherical
symmetry. However, we effectively identify the energy-momentum
tensors as
\begin{equation}\label{efective T}
    {T^{(\rm tot)}}_{\mu}{}^{\nu}={\rm diag}[-\tilde{\epsilon},\tilde p_r,\tilde p_t,\tilde p_t],
\end{equation}
\begin{equation}\label{T diaog}
{T^{(\rm m)}}_{\mu}{}^{\nu}={\rm diag}[-\epsilon,p_r,p_t,p_t],
\end{equation}
\begin{equation}\label{t diaog}
    {T^{(\rm X)}}_{\mu}{}^{\nu}={\rm
    diag}[-\mathcal{E},\mathcal{P}_r,\mathcal{P}_t,\mathcal{P}_t],
\end{equation}
where ($\tilde{\epsilon}$, $\epsilon$ and $\mathcal{E}$) are
energy densities, ($\tilde p_r$, $p_r$ and $\mathcal{P}_r$) are
radial pressure densities, and ($\tilde p_t$, $p_t$ and
$\mathcal{P}_t$) are tangential pressure densities. In general,
when $\tilde{p_t}\neq\tilde{p_r}$, these definitions clearly
indicate an anisotropy case.

In addition, the conservation equation
\begin{equation}\label{coservation eq}
    \nabla_{\mu}T^{({\rm tot})\,\mu\nu}=0,
\end{equation}
with metric \eqref{line elemets1}, is a linear combination of
Eqs.~\eqref{00}-\eqref{22}. However, in terms of those two sources
in relation~\eqref{ems}, while using relations \eqref{efective
T}-\eqref{t diaog}, it gives
\begin{equation}\label{coservation eq1}
 \begin{split}
&\frac{2}{r}\left(p_r-p_t+\mathcal{P}_r-\mathcal{P}_t\right)+\frac{\nu'}{2}
\left(\epsilon+p_r+\mathcal{E}+\mathcal{P}_r\right)\\&+\left(p'_r+\mathcal{P}'_r\right)=0.
\end{split}
\end{equation}

Now, to solve the system Eqs.~\eqref{00}-\eqref{22}, we implement
the GD method, which is explained in detail in Ref.~\cite{extended
MGD}, of course, we mention the necessary steps very briefly. We
assume that the solution of Eqs.~\eqref{00}-\eqref{22} for the
source $T^{(\rm m)}_{\mu\nu} $ when $T^{(\rm X)}_{\mu\nu}=0$ is a
general static spherically symmetric one as
\begin{equation}\label{line elementt=0}
    ds^2|_{T^{(\rm X)}_{\mu\nu}=0}\!=-e^{\zeta(r)}dt^2+e^{\lambda(r)}dr^2+r^2(d\theta^2+sin^2{\theta}\,
    d\phi^2).
\end{equation}
The energy-momentum tensor $T^{(\rm m)}_{\mu\nu} $ is conserved
with this metric, i.e.
\begin{equation}\label{coservationEqm}
    \widetilde{\nabla}_{\mu}T^{({\rm m})\,\mu\nu}=0,
\end{equation}
where $\widetilde{\nabla}_{\mu}$ is calculated according to metric
\eqref{line elementt=0}. We also assume that the effects of the
presence of the source $T^{(\rm X)}_{\mu\nu}$ on solution
\eqref{line elementt=0} are in the geometric deformation
\begin{equation}\label{deformation1}
\zeta\to\nu=\zeta+\alpha\, g(r)
\end{equation}
and
\begin{equation}\label{deformation2}
e^{-\lambda}\to e^{-\mu}=e^{-\lambda}+\alpha\, f(r),
\end{equation}
where $f(r)$ and $g(r)$ are geometric deformations that alter the
radial and temporal metric components, respectively, and the
constant $\alpha$ is a free positive parameter, which can somehow
adjust the strength of the effects on these components
simultaneously. Moreover, having $\alpha$ guarantees that
solution~\eqref{line elementt=0} is recovered in the limit
$\alpha\to 0$. We refer to the parameter $\alpha$ as a deformation
parameter.

Then, substituting decompositions~\eqref{deformation1}
and~\eqref{deformation2} into the original field
Eqs.~\eqref{00}-\eqref{22} causes those to be separated into two
decoupled sets of equations. One set of the field equations is for
the standard energy-momentum tensor $T^{(\rm m)}_{\mu\nu}$, i.e.
\begin{equation}\label{00T}
    \kappa\,\epsilon=-\Lambda -e^{-\lambda } \left(\frac{1}{r^2}-\frac{\lambda
    '}{r}\right)+\frac{1}{r^2},
\end{equation}
\begin{equation}\label{11T}
    \kappa\, p_r=\Lambda +e^{-\lambda } \left(\frac{1}{r^2}+\frac{\zeta
    '}{r}\right)-\frac{1}{r^2},
\end{equation}
\begin{equation}\label{22T}
   \kappa\, p_t=\Lambda +\frac{1}{4} e^{-\lambda } \left[-\lambda ' \zeta '+2 \zeta ''
   +\zeta '^2+\frac{2 \left(\zeta '-\lambda '\right)}{r}\right].
\end{equation}
Another set of the field equations is for the source $T^{(\rm
X)}_{\mu\nu}$, i.e.
\begin{equation}\label{00teta}
    \kappa\,\mathcal{E}=-\frac{\alpha f}{r^2}-\frac{\alpha f'}{r},
\end{equation}
\begin{equation}\label{11teta}
    \kappa\, \mathcal{P}_r-\alpha \frac{e^{-\lambda} g'}{r}=\alpha f
    \left(\frac{1}{r^2}+\frac{\nu'}{r}\right),
\end{equation}
\begin{eqnarray}\label{22teta}
 &&\kappa\,\mathcal{P}_t-\frac{\alpha e^{-\lambda}}{4}\left(2 g''+\alpha g'^2+\frac{2 g'}{r}+2 g' \zeta'-\lambda' g'\right) =\cr
    &&\frac{\alpha f}{4}\left(2 \nu''+\nu'^2+2\frac{\nu'}{r}\right)
    +\frac{\alpha f'}{4}\left(\nu'+\frac{2}{r}\right),
\end{eqnarray}
which clearly shows that when the deformation parameter is zero,
this source vanishes.

At this stage, one can investigate the field
Eqs.~\eqref{00T}-\eqref{22T} to determine~(${T^{(\rm m)}_{\mu\nu},
\zeta, \lambda}$), and then solve the field
Eqs.~\eqref{00teta}-\eqref{22teta} to specify~($ {T^{(\rm
X)}_{\mu\nu}, g, f }$). In other words, both sets, although
separated, remain gravitationally connected, i.e. to solve the
second set, one needs first to solve the first set. However, since
the number of unknowns in each set of equations is more than the
number of independent equations, it is necessary to apply
additional conditions and/or extra relations, e.g. the equation of
state. More explicitly, after solving the first set, five unknowns
remain in the three equations corresponding to the additional
source. Hence, to solve this set, we need to impose two
constraints, which we will perform in the next section.

\section{Deformation of AdS-Schwarzschild Black Hole }
In order to deform AdS-Schwarzschild black hole for action
\eqref{main action}, we start in the absence of $L_{\rm X}$ and
with the solution of the exterior Schwarzschild (i.e., in vacuum
with $T^{(\rm m)}_{\mu\nu} = 0$) in the AdS background. In this
regard, the field Eqs.~\eqref{enesteineq} give
\begin{equation}\label{Ads}
    e^{\zeta(r)}|_{T^{(\rm m)}_{\mu\nu}=0}=e^{-\lambda(r)}|_{T^{(\rm m)}_{\mu\nu}=0}=1-\frac{2M}{r}+\frac{r^2}{l^2}
\end{equation}
for the region $r > R$, where $R$ is the surface of the
self-gravitating system, $M$ is the ADM mass and
$l=\sqrt{3/|\Lambda|}$ is the AdS radius.

In order to have black holes with a well-defined event horizon
structure for action \eqref{main action}, the sufficient condition
(i.e., the Kerr-Schild condition)
\begin{equation}\label{hcondition}
    e^{\mu (r)}=e^{-\nu (r)}
\end{equation}
can be imposed~\cite{Ovalle2023} on metric \eqref{line elemets1}.
Actually, by adding this ansatz, the deformed black hole metric is
a metric that respects the symmetries. Accordingly, a direct
consequence of Eqs.~\eqref{00} and \eqref{11} plus
condition~\eqref{hcondition} is that the total source $T^{(\rm
tot)}_{\mu\nu}$ must satisfy the equation of state
\begin{equation}\label{state2}
\tilde{p_r}=-\tilde{\epsilon}.
\end{equation}
When $T^{(\rm m)}_{\mu\nu} = 0$, this relation yields the
constraint
\begin{equation}\label{state1}
\left[\mathcal{P}_r=-\mathcal{E}\right]\big|_{T^{(\rm m)}_{\mu\nu}
= 0},
\end{equation}
i.e., in this case, with positive energy density of the source
$T^{(\rm X)}_{\mu\nu}$, only its negative radial pressure density
is allowed. Furthermore, by deriving the conservation of the
energy-momentum tensor of sources with metric~\eqref{line
elemets1} while using decompositions~\eqref{deformation1}
and~\eqref{deformation2} as well as using relations \eqref{T
diaog} and \eqref{coservationEqm}, the first equality in relation
\eqref{Ads}, and condition \eqref{hcondition}, gives\footnote{It
is instructive to emphasize that the use of the first equality in
relation \eqref{Ads} and condition \eqref{hcondition} is necessary
to obtain the result of Eq. \eqref{ExchangeEq}. }
\begin{equation}\label{ExchangeEq}
    \nabla_{\mu}T^{({\rm m})\,\mu}{}_{\nu}=-\frac{\alpha\, g'}{2}(\epsilon
    +p_r)\delta_{1\,\nu}
    =-\nabla_{\mu}T^{({\rm X})\,\mu}{}_{\nu}.
\end{equation}
This relation, while assures the decoupling of the field
equations, shows the exchange of energy between the sources.
Moreover, from relation~\eqref{ExchangeEq}, it is obvious that
there is no energy exchange in the absence of $T^{({\rm
m})}_{\mu\nu}$, or in the special case where $\epsilon =-p_r$,
and/or when the geometric deformation $g$ is constant.

From now on, for simplicity, we consider the background geometry
to be the AdS-Schwarzschild vacuum, i.e. when $T^{(\rm
m)}_{\mu\nu} = 0$ and we have solution~\eqref{Ads}, but henceforth
we do~not mention the subscript. In fact, in the continuation, our
aim is to explore the back-reaction of a static,
spherically-symmetric energy-momentum tensor on the
four-dimensional AdS-Schwarzschild black hole. Of course, it
should be noted that in the mentioned case, one can get the
solution without resorting to the GD method. However, we utilize
the GD method as an alternative approach to solve this case to
better observe the effect of moving away from the
AdS-Schwarzschild solution and also to benefit from varying the
deformation parameter when studying the thermodynamic properties.
In other words, the merit of the GD method, in general, is that it
is easy any interpretation in terms of energy exchange as well as
in terms of superposition of configurations.

Accordingly, in the next step, by substituting
condition~\eqref{hcondition} and the AdS-Schwarzschild
solution~\eqref{Ads} into decompositions~\eqref{deformation1}
and~\eqref{deformation2}, we obtain
\begin{equation}\label{con1}
\alpha\,
f(r)=\left(1-\frac{2M}{r}+\frac{r^2}{l^2}\right)[e^{\alpha\,
g(r)}-1].
\end{equation}
Hence, the line element \eqref{line elemets1} becomes
\begin{equation}\label{line elemets2}
    ds^2\! =\! - e^{\zeta(r)} B(r)dt^2+\frac{1}{e^{\zeta(r)} B(r)}dr^2+r^2(d\theta^2+sin^2{\theta}
    d\phi^2),
\end{equation}
where
\begin{equation}\label{def2}
B(r)\equiv e^{\alpha\, g(r)}.
\end{equation}
Therefore, to completely specify the line element \eqref{line
elemets1} in this case, we need to specify the behavior of
function $B(r)$ (or indeed $g(r)$).

Now, using Eqs.~\eqref{00teta} and \eqref{con1}, the differential
equation governing the function $B(r)$ reads
\begin{equation}\label{mdeq}
r^2 \kappa\, \mathcal{E}=\left(2M-r-\frac{r^3}{l^2}\right)
B'(r)+\left(1+\frac{3r^2}{l^2}\right)[1- B(r)].
\end{equation}
Eq.~\eqref{mdeq} shows that the behavior of function $B(r)$
depends on the function of energy density $\mathcal{E}$, which in
turn depends on the choice of $T^{(\rm X)}_{\mu\nu}$ as the
additional source. To proceed, we resort to the weak energy
condition for the $T^{(\rm X)}_{\mu\nu}$, namely
\begin{equation}\label{WEC1}
\mathcal{E}\ge 0,
\end{equation}
\begin{equation}\label{WEC2}
\mathcal{E}+\mathcal{P}_r\ge 0,
\end{equation}
\begin{equation}\label{WEC3}
\mathcal{E}+\mathcal{P}_t\ge 0.
\end{equation}
In the special case under consideration, due to
constraint~\eqref{state1}, condition~\eqref{WEC2} holds.
Considering Eq.~\eqref{mdeq}, condition~\eqref{WEC1} puts the
constraint
\begin{equation}\label{constraint}
\left(2M-r-\frac{r^3}{l^2}\right)
B'(r)+\left(1+\frac{3r^2}{l^2}\right)[1- B(r)]\ge 0
\end{equation}
on the function $B(r)$. While using the conservation
Eq.~\eqref{coservation eq1} plus constraint~\eqref{state1},
condition~\eqref{WEC3} imposes the constraint
\begin{equation}\label{WEC33}
    2(\mathcal{E}+\mathcal{P}_t)=-r \mathcal{E}'\ge 0
\end{equation}
on the function $\mathcal{E}(r)$.

Once again, to continue the process, we deliberately choose the
energy density function of the additional source as the special
monotonic function\footnote{Also, this function is a reminder that
the energy density of radiation is proportional to $1/r^4 $. }
\begin{equation}\label{Efunction}
    \mathcal{E}(r)=\frac{\alpha }{\kappa  (\beta +r)^4},
\end{equation}
which satisfies\footnote{Note that, the dimension of the
deformation parameter is the square of length in the natural
units.}\
 both condition~\eqref{WEC1} and
constraint~\eqref{WEC33}, and has rapid asymptotic decay. In fact,
considering Eq.~\eqref{00teta} and choosing
function~\eqref{Efunction}, we have selected a special function
for the geometric deformation $f(r)$ and in turn the function
$B(r)$. In function~\eqref{Efunction}, $r\neq -\beta$, where
$\beta$ is a constant parameter with length dimension that
controls the behavior of $\mathcal{E}(r)$ at $r=0$, i.e. it is
necessary to avoid the central singularity. We refer to the
parameter $\beta $ as a control parameter. In addition, the role
of the deformation parameter in function~\eqref{Efunction} is
consistent with Eq.~\eqref{00teta}. Meanwhile,
substituting~\eqref{Efunction} into Eq.~\eqref{WEC33} gives
\begin{equation}\label{presu-t}
  \mathcal{P}_t=\frac{r-\beta}{\beta+r}\mathcal{E},
\end{equation}
and in turn, by using relation \eqref{t diaog} and constraint
\eqref{state1}, the trace of the energy-momentum tensor $T^{(\rm
X)}_{\mu\nu} $ is obtained to be
\begin{equation}\label{traceTX}
  T^{(\rm
X)}=-\frac{4\beta}{\beta+r}\mathcal{E}.
\end{equation}
Thus, under the mentioned considerations, the energy-momentum
tensor $T^{(\rm X)}_{\mu\nu} $ depends only on a single function
of the radial coordinate, its energy density function.

Now, by substituting function~\eqref{Efunction} into
Eq.~\eqref{mdeq} and solving the differential equation for the
function $B(r)$, we obtain
\begin{equation}\label{Bsolve}
 B(r)=\frac{\frac{\alpha  \beta ^2 }{3 (\beta
+r)^3}+\frac{\alpha } {\beta +r}-\frac{\alpha  \beta }{(\beta
+r)^2}+ r+\frac{r^3+c_1}{l^2}}{ r(1-2 M/r+r^2/l^2)},
\end{equation}
where $c_1$ is an integration constant. However, to have the
background AdS-Schwarzschild solution~\eqref{Ads} in the limit $
\alpha\to 0$ (i.e., when $B(r){\xrightarrow[\alpha\to
0]{\hspace*{0.6cm}}} 1$, and consistent with definition
\eqref{def2}), it is necessary to set\footnote{Note that, due to
function \eqref{Efunction}, solution \eqref{Bsolve} with value
\eqref{cvalue} automatically satisfies constraint
\eqref{constraint}.}
\begin{equation}\label{cvalue}
c_1=- 2 M l^2.
\end{equation}
Hence, the metric function reads
\begin{equation}\label{metric function}
  e^{\zeta(r)} B(r)=1-\frac{2M}{r}+\frac{r^2}{l^2} +\alpha\,\frac{\beta^2+3r^2+3\beta r}{3r(\beta+r)^3}\equiv
  F(r),
\end{equation}
which preserves the asymptotic behavior in the form of the AdS
solution. Moreover, the asymptotic behavior of the term resulting
from the additional source is $1/r^2$, i.e. like a Maxwell field.

Therefore, the line element for the region outside a
self-gravitating system filled with the energy-momentum tensor
$T^{(\rm X)}_{\mu\nu} $, while choosing the behavior of its energy
density as function~\eqref{Efunction}, is
\begin{equation}\label{final line element}
    ds^2=-  F(r)~dt^2+\frac{1}{F(r)} ~dr^2+r^2(d\theta^2+sin^2{\theta}\,
    d\phi^2).
\end{equation}
We refer to the black hole solution of this metric as a deformed
AdS-Schwarzschild black hole, and in the next section, investigate
its horizon structure and thermodynamics.

\section{ Structure of Horizon and Thermodynamics of Deformed AdS-Schwarzschild Black Hole  }
First of all, the metric function~\eqref{metric function} in the
limit $\beta\to 0$ becomes the metric function corresponding to
the well-known charged AdS black hole, i.e.
\begin{equation}\label{metric function for beta=0}
    F(r)|_{\beta\to 0}= 1-\frac{2M}{r}+\frac{r^2}{l^2}+\frac{\alpha}{r^2}.
\end{equation}
Therefore, $\sqrt{\alpha}$ can be interpreted as the electric
charge of black hole. In this particular case, the thermodynamic
behavior of the black hole is reminiscent of that investigated in
Refs.~\cite{Black hole chemistry,Stability and Hawking-Page-like
phase transition}. However, in these references, with the electric
charge as a conserved charge of the theory, the possibility of
variable electrostatic energy with an electric potential has been
considered, while we consider the deformation parameter only as a
free parameter\rlap.\footnote{Of course, the deformation parameter
can also be considered as an additional hair that is~not related
to other charges, i.e. the mass, the electric charge and the
angular-momentum.}\
 Moreover, we emphasize that in the case of
$\beta =0$, relation~\eqref{traceTX} indicates that the trace of
the additional energy-momentum tensor is zero, and utilizing
constraint~\eqref{state1} and relation~\eqref{presu-t} gives
$\mathcal{P}_{\rm eff}\equiv
\frac{1}{3}\(\mathcal{P}_r+\mathcal{P}_t+\mathcal{P}_t\)=\frac{1}{3}\mathcal{E}$.
These results plus the second footnote confirm that the additional
energy-momentum tensor is the one of a Maxwell field and its
solution is simply the electrically charged static AdS black hole.

On the other hand, we deliberately put the control parameter, i.e.
$\beta $, by hand in function~\eqref{Efunction} so that $\beta\neq
0$ prevents the divergence of the energy density~$\mathcal{E}$.
Therefore, it is better to check the effect of the control
parameter on the behavior of the metric function. For this
purpose, we expand the metric function~\eqref{metric function}
around the zero of its argument to obtain
\begin{equation}\label{metric function aroundr=0}
    F(r)|_{r\to 0}= \frac{\frac{\alpha }{\beta }-6 M}{3 r}+1+\frac{1}{3}
    r^2 \left(\frac{3}{l^2}-\frac{\alpha }{\beta ^4}\right)+{\cal
O}\left(r^3\right).
\end{equation}
Hence, with the value of
\begin{equation}\label{not divrge}
    \beta=\frac{\alpha}{6 M},
\end{equation}
the metric function $ F(r) $ has no singularity at $r = 0$.
Further investigation of this particular case may lead to a
regular black hole, however we leave this investigation to another
work. That is, although relation~\eqref{not divrge} is necessary
to avoid the divergence of the metric function, we do~not limit
our investigation to it in this work.

However, by the simple transformation $r\rightarrow \beta+r$, the
control parameter is turned on. Hence, the energy-momentum tensor
$T^{(\rm X)}_{\mu\nu} $ no longer describes a pure Maxwell field
and the corresponding metric is modified. Actually, we are
interested to have the black hole solution of metric \eqref{final
line element} in the presence of non-vanishing $\beta $. For this
purpose, the equation governing the radius of the event horizon
$r_{\rm h}$ is determined by the larger root of equation
\begin{equation}\label{horizon }
    F(r_{\rm h})=0,
\end{equation}
which is a sixth order equation. Its solution is complicated and
not useful for our purpose. Hence, instead of solving it
analytically, we restrict ourselves to specifying its important
consequences by resorting to its various corresponding diagrams in
FIG.~$1$ with fixed values of its constants.
\begin{figure*}[t!] \label{ffig1}
    \centering { \includegraphics[width=0.7\textwidth]{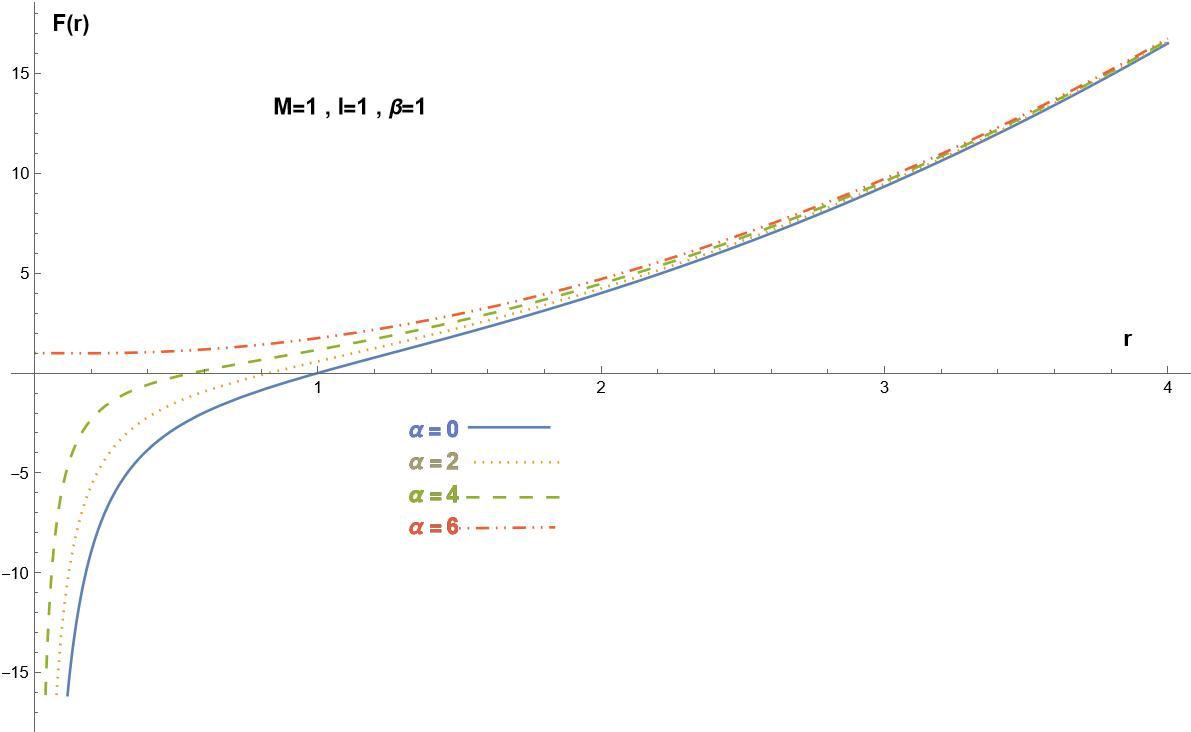} }
\label{fig1}\caption{[color online] Using relation \eqref{metric
function}, the figure schematically (i.e., scale-free) illustrates
the behavior of the function $F(r)$ with respect to $r$ for fixed
values of $ M=1$, $l=1$ and $\beta=1$, and different chosen values
of $\alpha$. As $\alpha$ increases, the radius of the event
horizon decreases, until $\alpha_{\rm max}\approx 6 $.}
\end{figure*}
For each of the curves in this figure, the intersection of the
metric function with the horizontal axis determines its
corresponding event horizon radius. In this regard, this figure
shows that the condition of having an event horizon limits the
value of the deformation parameter to an upper bound, $\alpha_{\rm
max}$. In general, the value of $\alpha_{\rm max}$ depends on the
values of other parameters, namely $M$, $l$ and $\beta$. For
instance, FIG.~$ 1$ indicates that when $M=1$, $l=1$ and
$\beta=1$, the radius of the event horizon decreases with
increasing the value of $\alpha$, until we
obtain\footnote{However, using Eq. \eqref{horizon } with these
mentioned values gives $\alpha =3(2-r_{\rm h}-r^2_{\rm
h})(1+r^3_{\rm h})/(1+3r_{\rm h}+3r^2_{\rm h})$ that yields
$\alpha { \xrightarrow[r_{\rm h}\to 0]{\hspace*{0.7cm}}} 6$.}\
 $\alpha_{\rm max}\approx 6 $. In the
continuation of this work, we pay attention to this limitation. In
other words, we comply the value $0<\alpha<\alpha_{\rm max}$ as a
condition for having an event horizon for the deformed
AdS-Schwarzschild black hole.
\begin{figure*}[t!] \label{ffig2}
    \centering {\includegraphics[width=0.7\textwidth]{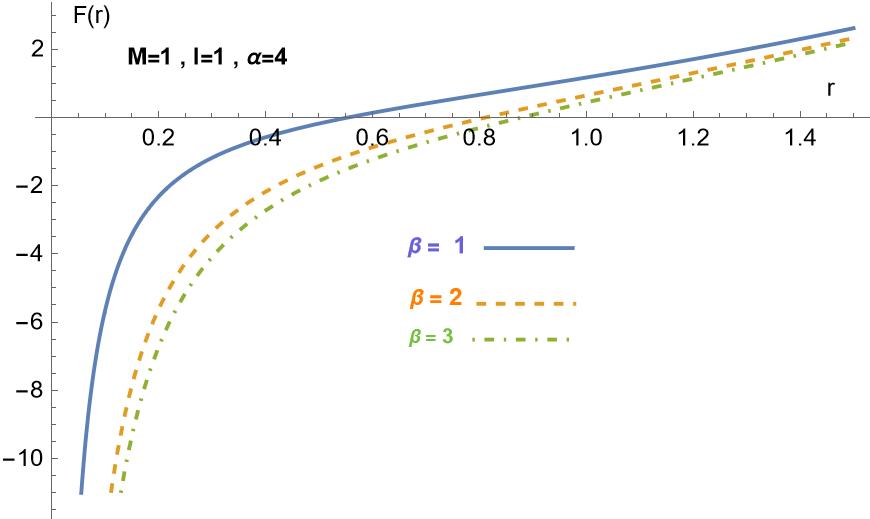} }
\label{fig2}\caption{[color online] Using relation \eqref{metric
function}, the figure schematically (i.e., scale-free) illustrates
the behavior of the function $F(r)$ with respect to $r$ for fixed
values of $ M=1$, $l=1$ and $\alpha=4$, and different chosen
values of $\beta$. As $\beta$ increases, the radius of the event
horizon increases.}
\end{figure*}

Alternatively, we have plotted the metric function $F(r)$,
relation~\eqref{metric function}, with respect to $r$, while
varying the control parameter $\beta $ and fixing the other
parameters including the deformation parameter $\alpha $ in
FIG.~$2$. This figure shows that when $M=1$, $l=1$ and $\alpha=4$,
the radius of the event horizon increases with increasing the
value of $\beta$. This effect is expected because the $\beta $
parameter shifts the radial coordinate as mentioned earlier. In
turn, increasing the radius of the event horizon affects the
thermodynamic properties, and this indicates the influence of the
control parameter on them.

To investigate the thermodynamic behavior of the obtained
solution, while using Eq.~\eqref{horizon }, we first express the
mass of the black hole in terms of the radius of the event
horizon, i.e.
\begin{equation}\label{Mass in term of r}
    M=\frac{1}{6} \left[3 r_{\rm h}+\frac{3 r_{\rm h}^3}{l^2}+\alpha\frac{
    \beta ^2+3 r_{\rm h}^2+3 \beta  r_{\rm h}}{(\beta +r_{\rm h})^3}\right].
\end{equation}
Furthermore, the entropy of the deformed AdS-Schwarzschild black
hole can be obtained from the Bekenstein-Hawking
formula~\cite{BE,GH77} as a quarter of the area of the event
horizon, i.e.
\begin{equation}\label{BH entropy}
    S\equiv\frac{A_{\rm h}}{4 L^2_{\rm Pl}}= \pi r_{\rm h}^2,
\end{equation}
where the Planck length $L_{\rm Pl}$ is considered in the natural
units and the second equality is due to the static and spherical
symmetry. Also, the definition of black hole pressure (density)
$P$ in AdS space (in the natural units with a negative
cosmological constant, see, e.g., Ref.~\cite{Black hole
chemistry}) is
\begin{equation}\label{BH press}
    P=\frac{3 }{8 \pi l^2 }.
\end{equation}
Then by obtaining $r_{\rm h}$ and $l$ from relations \eqref{BH
entropy} and \eqref{BH press}, and substituting into
relation~\eqref{Mass in term of r}, the mass of the deformed
AdS-Schwarzschild black hole can also be expressed in terms of
thermodynamic quantities $S$ and $P$ as
\begin{equation}\label{Mass in term of thermo}
    M=\frac{1}{6} \left[3\sqrt{\frac{S}{\pi }}+8PS\sqrt{\frac{S}{\pi }}+\alpha\frac{
    \beta ^2+3\frac{ S}{\pi }+3\beta\sqrt{\frac{S}{\pi }}
    }{\left(\beta +\sqrt{\frac{S}{\pi }}\right)^3}\right].
\end{equation}
Hence, for the deformed AdS-Schwarzschild black hole, we can write
the (generalized) first law of black hole thermodynamics in the
extended phase-space as\footnote{Note that, in this law, we do~not
consider the deformation parameter as a possibility of variable
electrostatic energy, see, e.g., Ref.~\cite{Avalos-etal}.
Nevertheless, if one puts a term like $+\Phi\,d\alpha $ in the
right hand of relation~\eqref{firs law }, due to partial
derivatives,
 relations~\eqref{Hawking temperaturein thermo}, \eqref{volum thermo}
 and~\eqref{heat capasity} will~not change, and by
 definition~\eqref{Gibss thermo}, relation~\eqref{Gibss radus} will~not change either.}
\begin{equation}\label{firs law }
dM=TdS+VdP,
\end{equation}
where $T$ and $V$ are the Hawking temperature and the
thermodynamic volume, respectively. Of course, in
relation~\eqref{firs law }, the presence of $VdP$ instead of $PdV$
indicates that the mass $M$ is the enthalpy of black hole instead
of the internal energy.

Now, using relation \eqref{Mass in term of thermo}, let us derive
these thermodynamic quantities. In this regard, the temperature
reads
\begin{equation}\label{Hawking temperaturein thermo}
    T=\left(\frac{\partial M}{\partial S}\right)_{\rm P }=\frac{1+8 P
    S-\frac{\pi\,
    \alpha\, S}{\left(\sqrt{\pi } \beta +\sqrt{S}\right)^4}}{4 \sqrt{\pi S }
    }.
\end{equation}
Of course, this relation can also be derived using the definition
of the Hawking temperature in terms of the radius of the event
horizon, i.e.
\begin{equation}\label{Hawking temperature definition}
    T\equiv\frac{1}{4\pi} F'(r)|_{r=r_{\rm h}}=\frac{1+r_{\rm h}^2 \left[\frac{3}{l^2}
    -\frac{\alpha }{(\beta +r_{\rm h})^4}\right]}{4 \pi  r_{\rm
    h}},
\end{equation}
where we have substituted $M$ from relation~\eqref{Mass in term of
thermo}. Then, taking $r_{\rm h}$ and $l$ from relations \eqref{BH
entropy} and \eqref{BH press}, and substituting into
relation~\eqref{Hawking temperature definition}, it reads the same
as relation~\eqref{Hawking temperaturein thermo} as expected.
Also, the thermodynamic volume is \cite{Dolan2011,Cvetic}
\begin{equation}\label{volum thermo}
    V=\left(\frac{\partial M}{\partial P}\right)_{\rm S}=\frac{4 S^{3/2}}{3 \sqrt{\pi }}
\end{equation}
that, in terms of the radius of the event horizon, obviously reads
\begin{equation}\label{BH volum}
    V=\frac{4}{3}\pi r_{\rm h}^3.
\end{equation}

To continue investigation of the thermodynamic behavior of the
deformed AdS-Schwarzschild black black hole, we apply the
thermodynamic machinery suggested in Ref.~\cite{Black hole
chemistry}. We assume that the black hole occurs in a canonical
ensemble. We also assume that each related extended phase-space
contains a fixed value of the deformation parameter. In this case,
the Gibbs free energy is\footnote{We do~not consider the
deformation parameter in this relation either.}
\begin{equation}\label{Gibss thermo}
G\equiv M-TS,
\end{equation}
which in terms of the radius of the event horizon becomes
\begin{equation}\label{Gibss radus}
G =\frac{r_{\rm h}}{4} -\frac{r_{\rm h}^3}{4 l^2} +\alpha \frac{ 9
r_{\rm h}^3+12 \beta r_{\rm h}^2+8\beta^2 r_{\rm h}+
2\beta^3}{12(\beta +r_{\rm h})^4}.
\end{equation}
In this regard, it is known that the condition of better thermal
equilibrium globally corresponds with more negative values of the
Gibbs free energy. Also, the criterion of the phase transition is
related to the vanishing Gibbs free energy of the black hole.

Another useful quantity in the thermodynamic study of a black hole
is the specific heat capacity at constant pressure, $C_{\rm P}$,
which determines the local thermodynamic stability of the black
hole. In the case at hand, it is
\begin{equation}\label{heat capasity}
    C_{\rm P}\equiv T \left(\frac{\partial S}{\partial T}\right)_{\rm
    P}=2\pi\frac{1+
 r_{\rm h}^2 \left[\frac{3}{l^2}
    -\frac{\alpha }{(\beta +r_{\rm h})^4}\right]}{\frac{3}{l^2}-\frac{1}{r_{\rm h}^2}
    -\frac{\alpha }{(\beta +r_{\rm h})^4}+\frac{4 \alpha  r_{\rm h}}{(\beta +r_{\rm
    h})^5}},
\end{equation}
where we have used the chain rule and relation \eqref{Hawking
temperature definition}. However, the local thermodynamic
stability corresponds to positive $C_{\rm P}$ values.

On the other hand, among the various black hole phase transitions,
one of the most significant is the Hawking-Page phase
transition~\cite{Hptransition}. This is a study of the
thermodynamics between an AdS-Schwarzschild black hole and the
thermal AdS space. In this respect, we consider the constant
values of $\beta=1$ and\footnote{Actually, considering relation
\eqref{BH press}, we investigate the thermodynamic behaviors at
constant pressure.}\
 $l=1 $ and choose several
constant values of the deformation parameter within its allowed
range to investigate its effects on the stability of the black
hole and particularly on the Hawking-Page phase transition.
\begin{figure*}[t!]\label{ffig2}
    \centering { \includegraphics[width=0.7\textwidth]{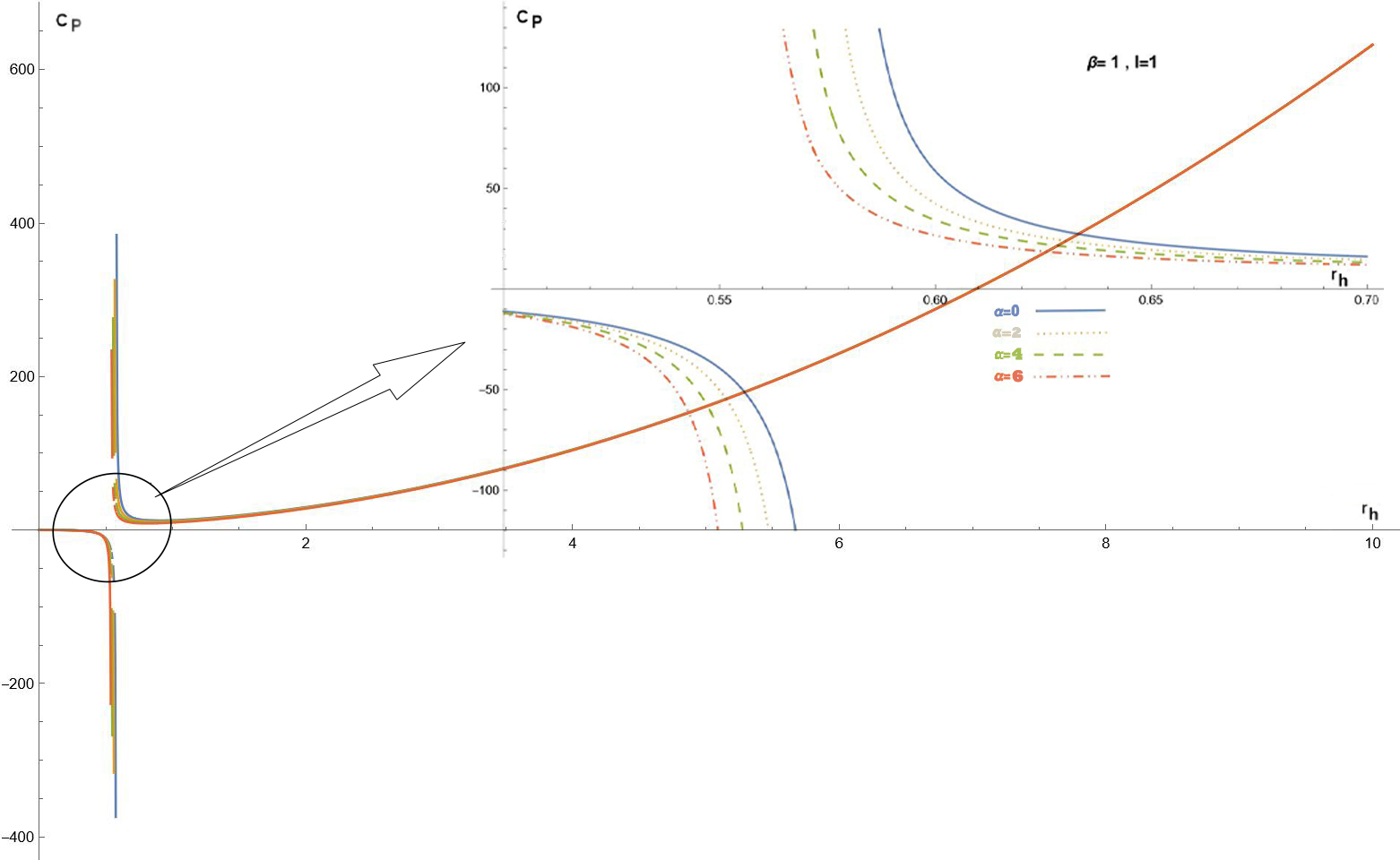} }
\label{fig2} \caption{[color online] Using relation \eqref{heat
capasity}, the figure schematically (i.e., scale-free) illustrates
the behavior of the $C_{\rm P}$ function with respect to $r_{\rm
h}$ for fixed values of $\beta=1$ and $l=1$, and different chosen
values of $\alpha$. In the $C_{\rm P}< 0$ region, the black holes
do~not have local thermodynamic equilibrium. In the $C_{\rm P}> 0$
region, as $\alpha$ increases, the black holes with smaller
horizon radii have local thermodynamic equilibrium. Also, in this
region for all curves, as $r_{\rm h}$ increases, the corresponding
value of $C_{\rm P}$ first decreases to a minimum and then
increases.}
\end{figure*}

First, let us investigate the thermodynamic stability of the black
hole locally through the behavior of the specific heat capacity at
constant pressure. In this regard, FIG.~$3$ indicates the behavior
of the $C_{\rm P}$ function with respect to the radius of the
event horizon for several values of the deformation parameter in
its allowed range. The resulting curves contain a discontinuity at
a certain horizon radius, such that for the chosen values it is
approximately in the range of~$0.55<r_{\rm h}<0.6$. Moreover, the
black holes with horizon radii located in the region of negative
$C_{\rm P}$ values are thermodynamically unstable, and a minimum
value of horizon radius is required for a black hole to have local
thermodynamic equilibrium. However, as the deformation parameter
increases, such minimum required radius decreases. FIG.~$3$ also
shows that black holes in the $C_{\rm P}< 0$ region have small
horizon radii (which we refer to as small black holes (SBHs)) and
transform into other thermodynamically allowed states. However,
black holes with larger horizon radii in the $C_{\rm P}>0$ region
(which we refer to as large black holes (LBHs)) have a clear local
thermodynamic stability. Hence, LBHs are thermodynamically
preferred over SBHs. In addition, according to this figure, as
$\alpha$ increases, the horizon radius of locally stable black
holes decreases. Furthermore, in the region of LBHs, for all
curves, as the horizon radius increases, the corresponding value
of $C_{\rm P}$ first decreases to a minimum and then increases. To
scrutinize the thermodynamic behavior of black holes globally, we
investigate the behavior of the Gibbs free energy function.
\begin{figure*}[t!]\label{ffig3}
    \centering { \includegraphics[width=0.7\textwidth]{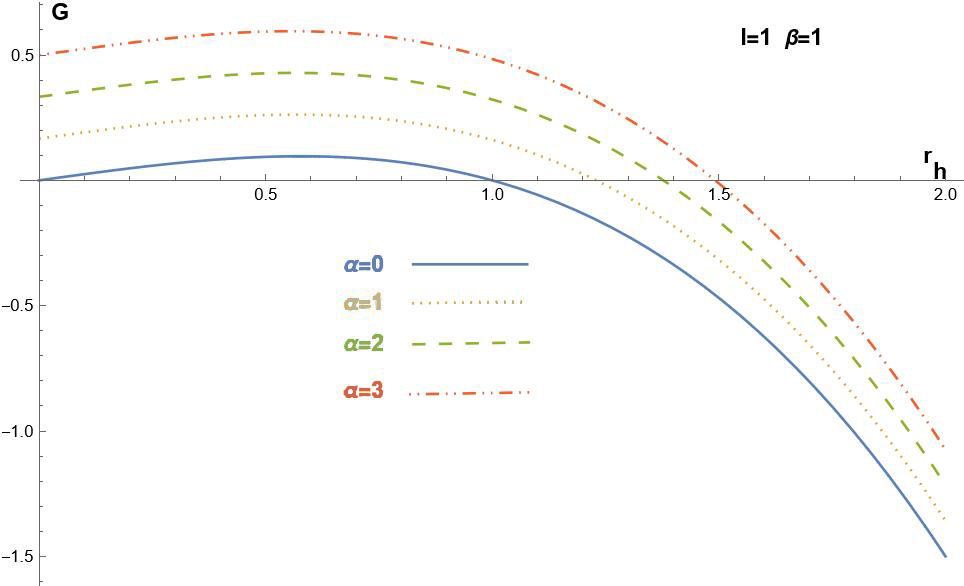} }
\label{fig3} \caption{[color online] Using relation \eqref{Gibss
radus}, the figure schematically (i.e., scale-free) illustrates
the behavior of the function $G$ with respect to $r_{\rm h}$ for
fixed values of $\beta=1$ and $l=1$, and different chosen values
of $\alpha$. The intersection points of the curves with the
horizontal axis (i.e., $G=0$) indicate the horizon radius of the
phase transition. As $\alpha$ increases, the $r_{\rm h}$ of the
phase transition and the maximum value of the $G$ function also
increase. More interestingly, in the $G< 0$ region, black holes
with larger $r_{\rm h}$ have less negative $G$ function.}
\end{figure*}

In this regard, first we have depicted the $G$ function versus
$r_{\rm h}$ with fixed values of $\beta=1$ and $l=1$, and
different chosen values of $\alpha$ in FIG.~$4$. It is clear that
the global thermodynamic equilibrium is better achieved with less
negative values of the $G$ function. Here, as an important aspect,
this figure illustrates that LBHs have less negative values of the
$G$ function. Hence, FIG.~$4$ confirms that in addition to local
thermodynamic stability, this group of black holes also has global
thermodynamic stability, while SBHs do~not. Thus once again, LBHs
are thermodynamically preferred over SBHs. This figure also shows
that the function $G$ is first ascending and then descending for
all allowed values of the deformation parameter. However, as the
deformation parameter increases, these maximum values of the $G$
function and also the horizon radius of the phase transition
(wherein $G=0$) increase. Nevertheless, increasing the deformation
parameter causes the thermodynamic stability of a black hole to be
disturbed compared to its previous position. Also, each maximum
point of the $G$ function represents a horizon radius and a
temperature. However, to better realize the phase transition and
the global thermodynamic stability of black holes, we probe the
behavior of the $G$ function versus the horizon temperature in the
following figure. Although it is better and instructive to first
plot the temperature versus the radius of the event horizon.
 \begin{figure*}[t!]\label{ffig4}
    \centering { \includegraphics[width=0.7\textwidth]{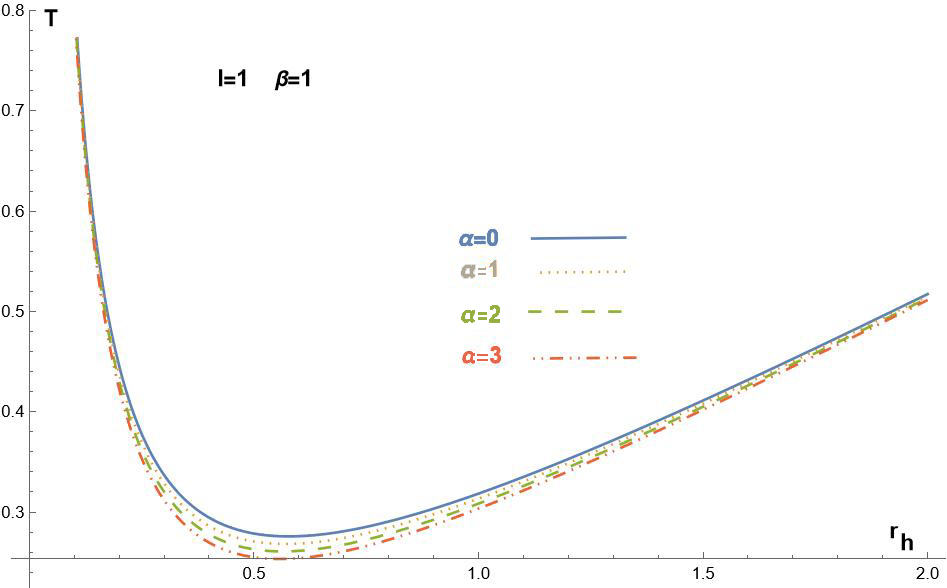} }
\label{fig4} \caption{[color online] Using relation \eqref{Hawking
temperature definition}, the figure schematically (i.e.,
scale-free) illustrates the behavior of the temperature of horizon
with respect to $r_{\rm h}$ for fixed values of $\beta=1$ and
$l=1$, and different chosen values of $\alpha$. As $r_{\rm h}$
increases, the horizon temperature first decreases to a minimum
and then increases. }
\end{figure*}

In this respect, FIG.~$5$ shows the behavior of the temperature of
horizon with respect to $r_{\rm h}$ for fixed values of $\beta=1$
and $l=1$, and different chosen values of $\alpha$. This figure
indicates that as $r_{\rm h}$ increases, the temperature of
horizon first decreases and then increases for all values of $
\alpha$. The minimum temperature in this figure exactly
corresponds to the maximum Gibbs function in FIG.~$4$. In general,
the changing behavior of the temperature function is in accord
with the behavior of the $G$ function.

Now, employing relations \eqref{Hawking temperature definition}
and \eqref{Gibss radus}, FIG.~$6$ shows the behavior of the Gibbs
function with respect to the horizon temperature for fixed values
of $\beta=1$ and $l=1$, and different chosen values of the
deformation parameter. The cusp of each curve in this figure
represents the maximum of the $G$ function (as shown in FIG.~$4$)
and the minimum temperature (as shown in FIG.~$5$). At
temperatures below the minimum temperature, there are no black
holes. FIG.~$6$ indicates that as the temperature increases from
its minimum, there are two branches of black holes. In both
branches the $G$ function decreases, which is in accordance with
FIG.~$4$. The right/upper branches contain SBHs with $C_{\rm
P}<0$, while the left/lower branches have $C_{\rm P}>0$ and
include black holes with intermediate horizon radii (which we
refer to as intermediate black holes (IBHs)) with positive values
of the $G$ function, the intersection of curves with the
horizontal axis (the phase transition with $G=0$), and LBHs with
negative values of the $G$ function.

Since thermodynamically, smaller (and even negative) values of the
$G$ function are preferred, there are actually two global
thermodynamically stable phases. The first phase in which $G=0$
represents the immersion environment of black holes and includes
the thermal radiation (the thermal bath filled with the
cosmological constant). This region is preferred with respect to
IBHs. The second phase belongs to LBHs, which indicate more
stability and thermodynamic preference. The proximity of the phase
transition between the thermal radiation medium and the deformed
AdS-Schwarzschild LBHs is known as the Hawking-Page phase
transition. The intersection of the curves with the horizontal
axis shows the temperature of the first-order Hawking-Page phase
transition. FIG.~$6$ also illustrates that as the deformation
parameter increases, the $G$ function and especially its maximum
increases (consistent with FIG.~$4$), the minimum temperature
decreases (consistent with FIG.~$5$) while the temperature of the
Hawking-Page phase transition increases.
\begin{figure*}[t!]\label{ffig5}
    \centering  { \includegraphics[width=0.7\textwidth]{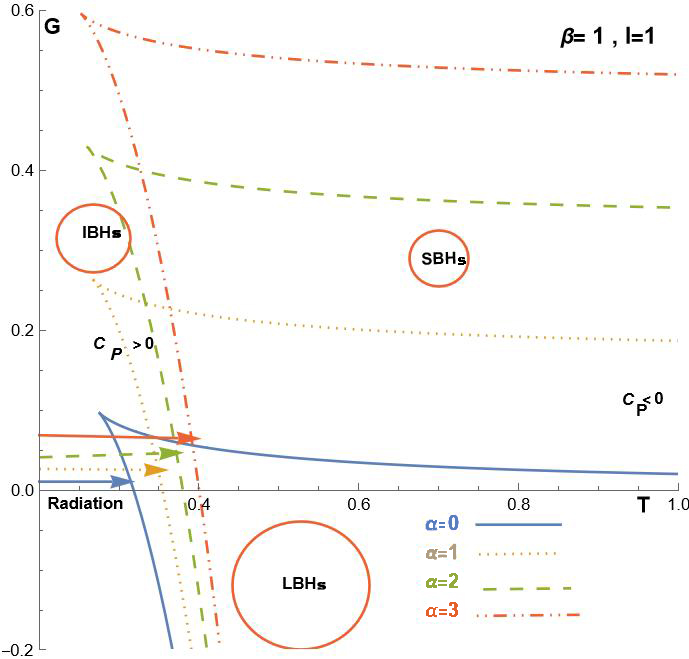} }
\label{fig5} \caption{[color online] The figure schematically
(i.e., scale-free) illustrates the behavior of the $G$ function
with respect to the horizon temperature for fixed values of
$\beta=1$ and $l=1$, and different chosen values of $\alpha$. A
deformed AdS-Schwarzschild black hole, like a Schwarzschild black
hole, exhibits a phase transition with a thermal radiation medium.
The region of the thermal radiation is indicated by an arrow for
each curve, and the end of each arrow indicates the temperature of
the first-order Hawking-Page phase transition. }
\end{figure*}

\section{ Conclusions }
Initially, we considered the Einstein-Hilbert action with the
presence of the cosmological constant and a standard matter source
plus any additional gravitational source. The aim of this work is
to find the black hole solution(s) for such an action and to
search for the corresponding thermodynamic behaviors. To proceed,
we have employed the GD method, which serves as a useful tool for
searching solutions to the gravitational equations. Then, we have
taken the background as the AdS-Schwarzschild vacuum solution, and
have looked for the static spherically symmetric solution(s) when
the additional source is also present. By implementing the GD
method, we have used two geometric deformation functions to alter
the radial and temporal components of the background metric.
Meanwhile, in process of this method, there is a common positive
parameter that can adjust the strength of the effects on these
components simultaneously, and we refer to it as a deformation
parameter. When this parameter vanishes, the background solution
is recovered. Through the GD method, the field equations are
separated into two decoupled sets of equations for each source,
although they remain gravitationally connected.

In continuation, after solving the set of equations for the
background, five unknowns remain in the three equations
corresponding to the additional source. Hence, to solve this set,
we imposed two constraints. Actually, we have assumed that the
additional source obeys the weak energy condition, and we have
also deliberately chosen its energy density function to be a
special monotonic function proportional to the inverse of the
distance to the fourth power. Also for consistency, this function
is proportional to the deformation parameter and includes a
control parameter to prevent it from diverging at the center.
Actually, in the absence of the control parameter, we showed that
the additional energy-momentum tensor is the one of a Maxwell
field. Moreover, to have black holes with a well-defined event
horizon structure, we have imposed the Kerr-Schild condition that
the radial and temporal components of the solution to be inverses
of each other regardless of their signs in the signature. We refer
to the obtained solution as a deformed AdS-Schwarzschild black
hole. The solution found, although complicated, turns out to be
analytical and with quite interesting features. Then, the focus of
the work has been to investigate the horizon structure, the
thermodynamics of the solution and the Hawking-Page phase
transition mainly by varying the deformation parameter. However,
within the work (including figures), we considered several values
of the deformation parameter with fixed values of other constants
(hence, actually at constant pressure).

Since we intended to consider the black hole solution, we first
plotted the metric function versus the distance. For each of the
curves in this figure, the intersection of the metric function
with the horizontal axis specifies that there is an event horizon
radius. However, the figure shows that the condition of having an
event horizon limits the value of the deformation parameter to an
upper bound, as we have also shown through the corresponding
derivation. Hence, we confined the investigation to vary the
deformation parameter up to its upper bound value.

Next, we assumed that the black hole occurs in a canonical
ensemble and wrote the first law of black hole thermodynamics in
the extended phase-space and then derived various thermodynamic
quantities as a function of the event horizon radius. We also
assumed that each related extended phase-space contains a fixed
value of the deformation parameter as a free parameter and not~a
thermodynamic quantity. Thereafter, to determine the thermodynamic
stability of the black hole locally, we plotted the obtained heat
capacity at constant pressure versus the radius of the event
horizon. The figure shows that a discontinuity occurs between the
negative and positive values of the heat capacity at constant
pressure for all its curves. The negative region contains SBHs
that are locally thermodynamically unstable. Whereas, the positive
region contains LBHs that are locally thermodynamically stable. In
other words, for a black hole to have local thermodynamic
equilibrium, a minimum value of horizon radius is required, and as
the deformation parameter increases, this minimum required radius
decreases. In addition, in the region of LBHs, for all curves, as
the horizon radius increases, the corresponding value of $C_{\rm
P}$ first decreases to a minimum and then increases, while as the
deformation parameter increases, the horizon radius of locally
stable black holes decreases.

Furthermore, to determine the thermodynamic stability of the black
hole globally, we plotted the obtained Gibbs free energy versus
the radius of the event horizon. This figure shows that the Gibbs
function starts from the region of positive values, and with the
increase of the horizon radius, it first increases to a maximum
and then decreases to more negative values after crossing the
horizontal axis (i.e., the phase transition point). The figure
also illustrates that LBHs have less negative values of the Gibbs
free energy. Hence, in addition to local thermodynamic stability,
this group of black holes also has global thermodynamic stability
and is thermodynamically preferred over SBHs. Moreover, with the
increase of the deformation parameter, the maximum values of the
Gibbs function and the horizon radius of the phase transition
increase. Nevertheless, increasing the deformation parameter
causes the thermodynamic stability of a black hole to be disturbed
compared to its previous position.

Then, to better realize the phase transition and global
thermodynamic stability of black holes, we plotted the temperature
versus the event horizon radius and the Gibbs free energy versus
the horizon temperature. The figures indicate that there is a
minimum temperature that exactly corresponds to the maximum of the
Gibbs function. As the deformation parameter increases, the
minimum temperature decreases and no~black hole exists below this
minimum temperature. The second figure illustrates that SBHs are
in one branch of the curves, and in the other branch are IBHs, the
phase transition point (i.e., thermal radiation) and LBHs. The
proximity of the phase transition between the thermal radiation
medium and the deformed AdS-Schwarzschild LBHs is known as the
Hawking-Page phase transition. As the deformation parameter
increases, the temperature of the first-order Hawking-Page phase
transition increases.

In the special case of vanishing the deformation parameter, the
obtained thermodynamic behavior of the black hole is consistent
with that stated in Ref.~\cite{Black hole chemistry}. Also, we
showed that in the special case of vanishing the control
parameter, the obtained metric function corresponds to the charged
AdS black hole, which was investigated in Refs.~\cite{Black hole
chemistry,Stability and Hawking-Page-like phase transition} with
the square of the deformation parameter as the role of electric
charge. In addition, we showed that increasing the control
parameter increases the radius of the event horizon, which in turn
affects the thermodynamic properties. Furthermore, we indicated
that for a certain non-zero value of the control parameter, the
obtained metric function has no singularity at the center. Further
investigation of this particular case may lead to a regular black
hole, however we leave these investigations to another work.

\section*{ACKNOWLEDGMENTS}
The authors thank the Research Council of Shahid Beheshti
University.


%
\end{document}